\documentclass[letterpaper,twocolumn,superscriptaddress,floatfix]{revtex4}
\usepackage{inputenc}
\usepackage{bm}
\usepackage{multirow,amssymb,amsbsy,amsmath}
\usepackage{graphicx}
\usepackage{float}
\usepackage{verbatim}
\makeatletter
\usepackage{pifont}
\usepackage{upgreek}
\usepackage{color}
\usepackage{indentfirst}
\usepackage{soul}
\usepackage[colorlinks,linkcolor=red,anchorcolor=blue,citecolor=green]{hyperref}

\begin{document}

\preprint{APS/123-QED}

\title{Temperature dependent single- and double-quantum relaxation of negatively charged boron vacancies in hexagonal boron nitride}

\affiliation{Laboratory of Quantum Information, University of Science and Technology of China, Hefei 230026, China}
\affiliation{Anhui Province Key Laboratory of Quantum Network, University of Science and Technology of China, Hefei 230026, China}
\affiliation{Department of Physics, School of Physics and Mechanics, Wuhan University of Technology, Wuhan 430070, China}
\affiliation{CAS Center For Excellence in Quantum Information and Quantum Physics, University of Science and Technology of China, Hefei 230026, China}
\affiliation{Hefei National Laboratory, University of Science and Technology of China, Hefei 230088, China}

\author{Lin-Ke Xie}
\thanks{These authors contributed equally to this work.}
\affiliation{Laboratory of Quantum Information, University of Science and Technology of China, Hefei 230026, China}
\affiliation{Anhui Province Key Laboratory of Quantum Network, University of Science and Technology of China, Hefei 230026, China}
\affiliation{CAS Center For Excellence in Quantum Information and Quantum Physics, University of Science and Technology of China, Hefei 230026, China}

\author{Wei Liu}
\thanks{These authors contributed equally to this work.}
\affiliation{Laboratory of Quantum Information, University of Science and Technology of China, Hefei 230026, China}
\affiliation{Anhui Province Key Laboratory of Quantum Network, University of Science and Technology of China, Hefei 230026, China}
\affiliation{CAS Center For Excellence in Quantum Information and Quantum Physics, University of Science and Technology of China, Hefei 230026, China}

\author{Kaiyu Huang}
\thanks{These authors contributed equally to this work.}
\affiliation{Department of Physics, School of Physics and Mechanics, Wuhan University of Technology, Wuhan 430070, China}

\author{Nai-Jie Guo}
\affiliation{Laboratory of Quantum Information, University of Science and Technology of China, Hefei 230026, China}
\affiliation{Anhui Province Key Laboratory of Quantum Network, University of Science and Technology of China, Hefei 230026, China}
\affiliation{CAS Center For Excellence in Quantum Information and Quantum Physics, University of Science and Technology of China, Hefei 230026, China}
\affiliation{Hefei National Laboratory, University of Science and Technology of China, Hefei 230088, China}

\author{Jun-You Liu}
\affiliation{Laboratory of Quantum Information, University of Science and Technology of China, Hefei 230026, China}
\affiliation{Anhui Province Key Laboratory of Quantum Network, University of Science and Technology of China, Hefei 230026, China}
\affiliation{CAS Center For Excellence in Quantum Information and Quantum Physics, University of Science and Technology of China, Hefei 230026, China}
\affiliation{Hefei National Laboratory, University of Science and Technology of China, Hefei 230088, China}

\author{Yu-Hang Ma}
\affiliation{Laboratory of Quantum Information, University of Science and Technology of China, Hefei 230026, China}
\affiliation{Anhui Province Key Laboratory of Quantum Network, University of Science and Technology of China, Hefei 230026, China}
\affiliation{CAS Center For Excellence in Quantum Information and Quantum Physics, University of Science and Technology of China, Hefei 230026, China}

\author{Ya-Qi Wu}
\affiliation{Laboratory of Quantum Information, University of Science and Technology of China, Hefei 230026, China}
\affiliation{Anhui Province Key Laboratory of Quantum Network, University of Science and Technology of China, Hefei 230026, China}
\affiliation{CAS Center For Excellence in Quantum Information and Quantum Physics, University of Science and Technology of China, Hefei 230026, China}

\author{Yi-Tao Wang}
\affiliation{Laboratory of Quantum Information, University of Science and Technology of China, Hefei 230026, China}
\affiliation{Anhui Province Key Laboratory of Quantum Network, University of Science and Technology of China, Hefei 230026, China}
\affiliation{CAS Center For Excellence in Quantum Information and Quantum Physics, University of Science and Technology of China, Hefei 230026, China}

\author{Zhao-An Wang}
\affiliation{Laboratory of Quantum Information, University of Science and Technology of China, Hefei 230026, China}
\affiliation{Anhui Province Key Laboratory of Quantum Network, University of Science and Technology of China, Hefei 230026, China}
\affiliation{CAS Center For Excellence in Quantum Information and Quantum Physics, University of Science and Technology of China, Hefei 230026, China}

\author{Xiao-Dong Zeng}
\affiliation{Laboratory of Quantum Information, University of Science and Technology of China, Hefei 230026, China}
\affiliation{Anhui Province Key Laboratory of Quantum Network, University of Science and Technology of China, Hefei 230026, China}
\affiliation{CAS Center For Excellence in Quantum Information and Quantum Physics, University of Science and Technology of China, Hefei 230026, China}

\author{Jia-Ming Ren}
\affiliation{Laboratory of Quantum Information, University of Science and Technology of China, Hefei 230026, China}
\affiliation{Anhui Province Key Laboratory of Quantum Network, University of Science and Technology of China, Hefei 230026, China}
\affiliation{CAS Center For Excellence in Quantum Information and Quantum Physics, University of Science and Technology of China, Hefei 230026, China}

\author{Chun Ao}
\affiliation{Laboratory of Quantum Information, University of Science and Technology of China, Hefei 230026, China}
\affiliation{Anhui Province Key Laboratory of Quantum Network, University of Science and Technology of China, Hefei 230026, China}
\affiliation{CAS Center For Excellence in Quantum Information and Quantum Physics, University of Science and Technology of China, Hefei 230026, China}

\author{Shuo Deng}
\affiliation{Department of Physics, School of Physics and Mechanics, Wuhan University of Technology, Wuhan 430070, China}

\author{Haifei Lu}
\altaffiliation{Email: haifeilv@whut.edu.cn}
\affiliation{Department of Physics, School of Physics and Mechanics, Wuhan University of Technology, Wuhan 430070, China}

\author{Jian-Shun Tang}
\altaffiliation{Email: tjs@ustc.edu.cn}
\affiliation{Laboratory of Quantum Information, University of Science and Technology of China, Hefei 230026, China}
\affiliation{Anhui Province Key Laboratory of Quantum Network, University of Science and Technology of China, Hefei 230026, China}
\affiliation{CAS Center For Excellence in Quantum Information and Quantum Physics, University of Science and Technology of China, Hefei 230026, China}
\affiliation{Hefei National Laboratory, University of Science and Technology of China, Hefei 230088, China}

\author{Chuan-Feng Li}
\altaffiliation{Email: cfli@ustc.edu.cn}
\affiliation{Laboratory of Quantum Information, University of Science and Technology of China, Hefei 230026, China}
\affiliation{Anhui Province Key Laboratory of Quantum Network, University of Science and Technology of China, Hefei 230026, China}
\affiliation{CAS Center For Excellence in Quantum Information and Quantum Physics, University of Science and Technology of China, Hefei 230026, China}
\affiliation{Hefei National Laboratory, University of Science and Technology of China, Hefei 230088, China}

\author{Guang-Can Guo}
\affiliation{Laboratory of Quantum Information, University of Science and Technology of China, Hefei 230026, China}
\affiliation{Anhui Province Key Laboratory of Quantum Network, University of Science and Technology of China, Hefei 230026, China}
\affiliation{CAS Center For Excellence in Quantum Information and Quantum Physics, University of Science and Technology of China, Hefei 230026, China}
\affiliation{Hefei National Laboratory, University of Science and Technology of China, Hefei 230088, China}

\date{\today}

\begin{abstract}
The negatively charged boron vacancy ($\rm V_B^-$) in two-dimensional (2D) hexagonal boron nitride (hBN) has emerged as a promising candidate for quantum sensing. 
The coherence time of $\rm V_B^-$ spins which coherent quantum sensing resides in is limited spin-phonon interactions, while the underlying physical mechanism of the corresponding high-temperature behavior is still not fully understood. 
Here, we probe the relaxation rates on $\rm V_B^-$ centers' $\left|m_s=0\right\rangle$ $\leftrightarrow$ $\left|m_s \pm 1\right\rangle$ (single-quantum) and $\left|m_s=+1\right\rangle$ $\leftrightarrow$ $\left|m_s=-1\right\rangle$ (double-quantum) over the temperature range from 293 to 393 K. 
The results show that both relaxation rates increase with increasing temperature, and the double-quantum relaxation rate significantly increases rapidly. At high temperature (above 400 K), the double-quantum relaxation rate is much greater than single-quantum relaxation rate, and may dominate the decoherence channel of spin-phonon interactions.
Using a theoretical model of second-order spin-phonon interactions, we attribute the high-temperature spin relaxation rates to interactions with higher-energy effective phonon mode, aiding the further understanding and guiding high-temperature sensing applications. 

\end{abstract}

\maketitle


\section{Introduction}

Solid-state spin defects have already shown huge application potential in quantum science and technology. To date, solid-state spin defects in diamond and silicon carbide have been extensively studied and widely used for their great stability and sensitivity under different probing conditions \cite{Kurtsiefer2000Stable,Jelezko2004Observation,Balasubramanian2009Ultralong,Herbschleb2019Ultralong,Koehl2011Room,Castelletto2014A,Widmann2015Coherent}. Though they have many exceptional features, such as excellent spin properties, fabrication and integrating them with other materials are challenging, especially when the defects must be close to the target sample \cite{Romach2015Spectroscopy}. Spin defects in two-dimensional (2D) van der Waals (vdW) crystals \cite{Liu2022Spin,Gottscholl2020Initialization,Mendelson2021Identifying,Chejanovsky2021Single,Stern2022Room,Guo2021Coherent,Yang2023Laser,Stern2024A,Scholten2024Multi,Liu2024Experimental} emerge as the excellent substitutes to address the problem, owing to the advantages of 2D layered structure. 2D hexagonal boron nitride (hBN) stands out as an excellent host for spin defects \cite{Liu2022Spin,Gottscholl2020Initialization,Mendelson2021Identifying,Chejanovsky2021Single,Stern2022Room,Guo2021Coherent,Yang2023Laser,Stern2024A,Scholten2024Multi}, which can be seamlessly integrated into heterostructures, optoelectronic and nanophotonic devices. Among a wide range of possible spin defects in hBN, the negatively charged boron vacancy ($\rm V_B^-$) is the most studied which can be optically initialized, coherently manipulated and read out \cite{Gottscholl2021Room,Liu2021Coherent}. Because $\rm V_B^-$ defect has a spin triplet ground state ($S=1$) \cite{Abdi2018Color,Ivady2020Ab} and is easy to create \cite{Gao2021Femtosecond,Guo2022Generation,Liang2023High}, it has been widely applied in sensing magnetic fields, temperature, strain, and so on \cite{Vaidya2023Quantum,Gottscholl2021Spin,Liu2021Temperature,Healey2021Quantum,Gao2023Quantum,Robertson2023Detection,Zeng2025Ambient}.

\begin{figure*}
\includegraphics[width=\textwidth,height=0.65\textwidth]{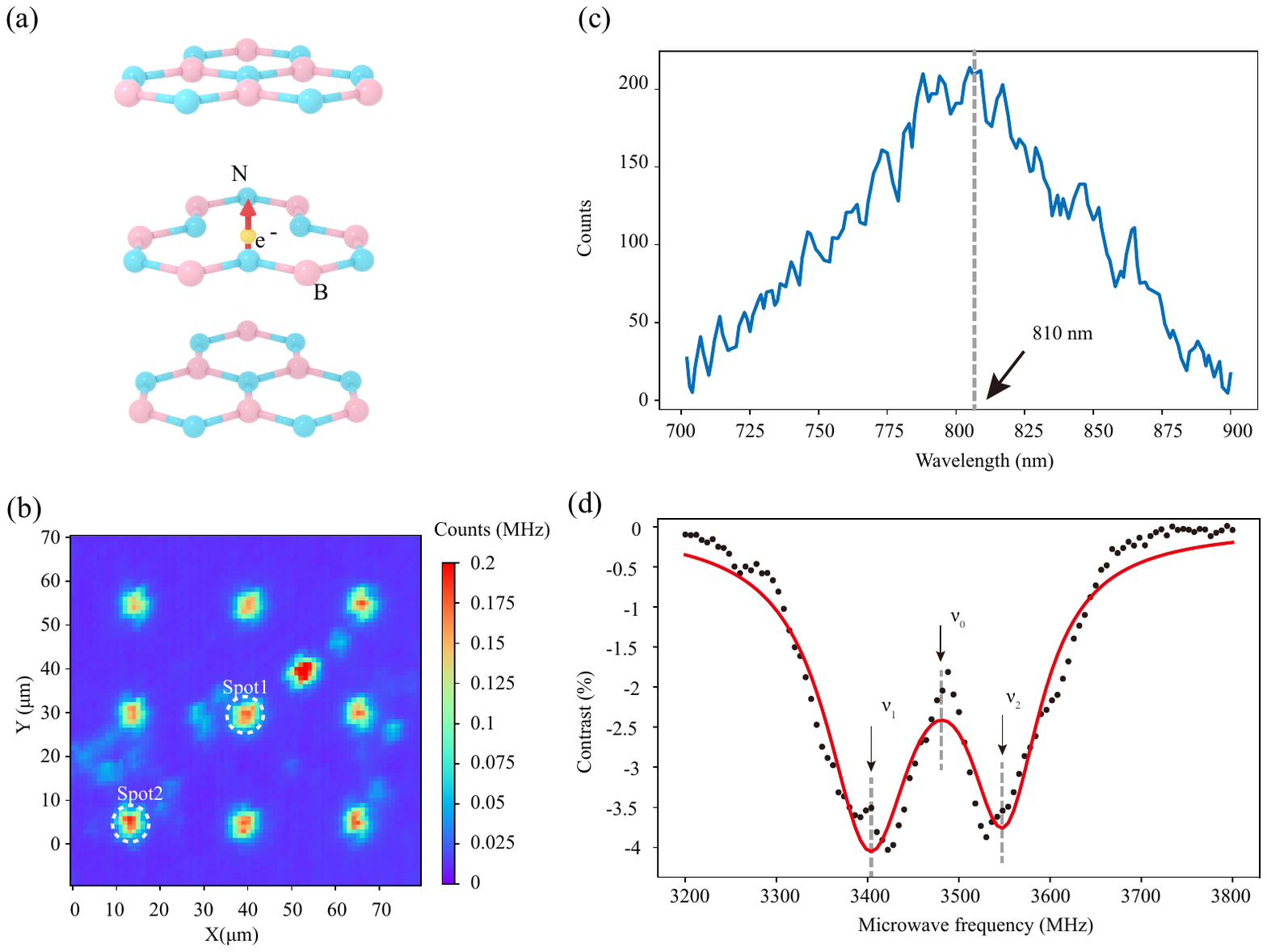}
\caption{\label{fig:wide} 
(a) Schematic of the $\rm V_B^-$ defect in hBN: an electron in boron vacancy (yellow) surrounded by three equivalent nitrogen atoms (blue). (b) Confocal PL intensity map of the flake hBN sample. Different colors in the colorbar correspond to photon counts per second. Areas circled by white dotted circles named Spot 1 and Spot 2 respectively, are the defect ensembles chosen for next study. (c)  PL spectrum of $\rm V_B^-$ ensemble defects under 532-nm laser excitation at room temperature. (d) ODMR spectrum measured at room temperature without external magnetic field, and the two-Lorentzian fits (red line) with the corresponding resonant frequencies  $\nu_1$ = 3401.82 $\pm$ 1.88 MHz, $\nu_2$ = 3549.27 $\pm$ 1.89 MHz and $\nu_0 = (\nu_1 +\nu_2)/2 = $ 3475.55 $\pm$ 1.89 MHz. } 
\end{figure*}

Interactions between the $\rm V_B^-$ defect's electronic spin and phonon in the surrounding crystal lattice drives spin-lattice relaxation, also called spin-phonon relaxation, play a crucial role in studying spin dynamics of the $\rm V_B^-$ defects and determine the upper limits of system's achievable electronic spin coherence time which coherent quantum sensing resides in \cite{Norambuena2018Spin,Xu2020Spin,Lunghi2022Toward}. Several works have already explored the temperature dependence of spin-lattice relaxation time $T_1$ corresponding to transitions between $\left|m_s = 0\right\rangle$ and $\left|m_s =\pm 1\right\rangle$ which is referred to as single-quantum relaxation \cite{Liu20225temperature}. However, transitions between $\left|m_s = -1\right\rangle$ and $\left|m_s = +1\right\rangle$ is termed the double-quantum relaxation rate, which remains an open question. 

In this letter, we fabricate an array of $\rm V_B^-$ defect ensembles by focused Helium ion beam in flake hBN, and use two designed measurement sequences to extract the single- and double-quantum relaxation rates. Applied a lab-built temperature control setup, the temperature dependence of the relaxation rates from 293 to 393 K is investigated. 
Using a theoretical model of second-order spin-phonon interactions \cite{Cambria2023Temperature} and the characteristic phonon modes of $\rm V_B^-$ obtained from first-principle calculations, we quantitatively analyze the characteristics of single- and double quantum relaxation processes. 

\begin{figure*}
\includegraphics[width=\textwidth,height=0.65\textwidth]{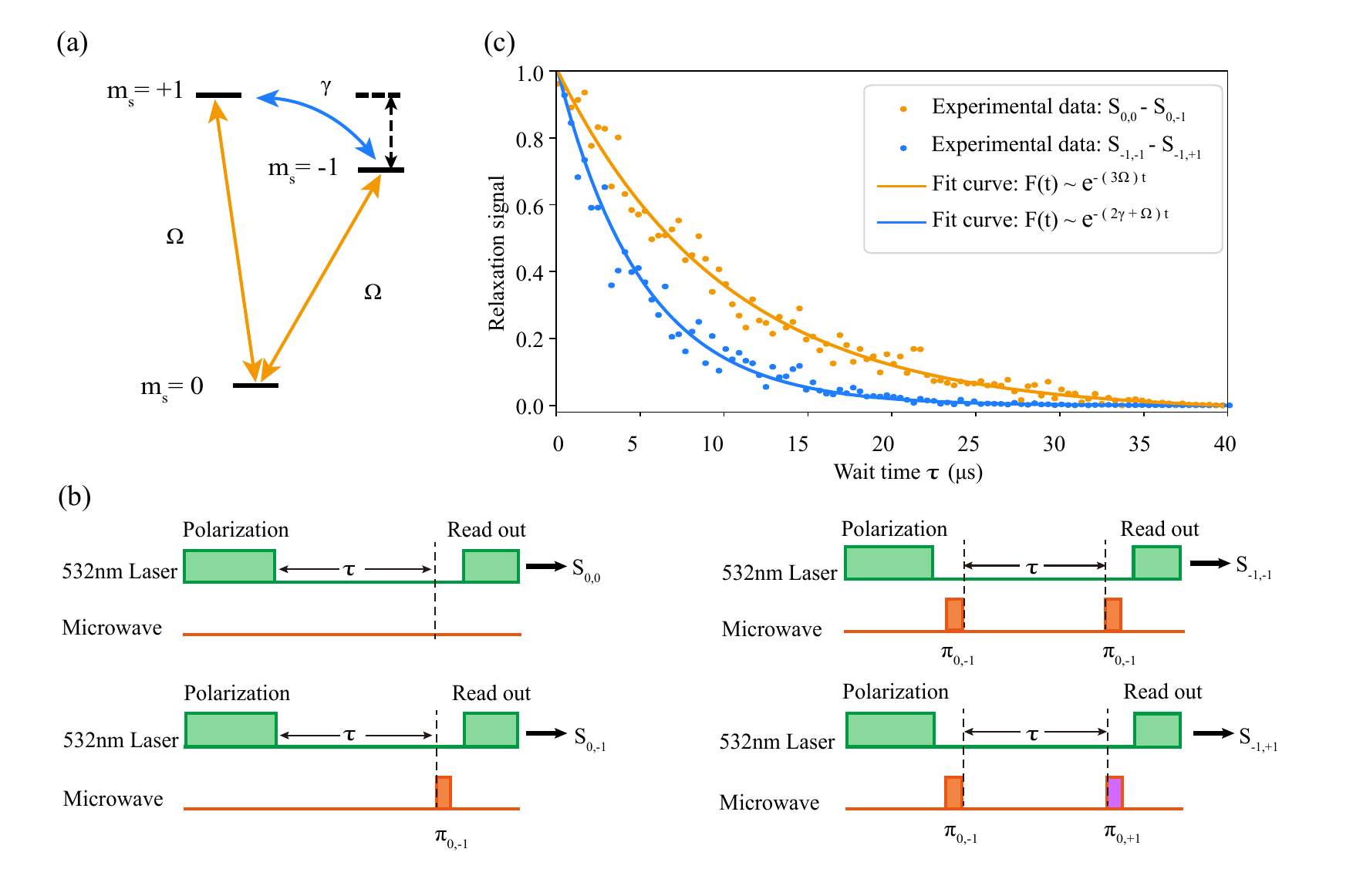}
\caption{\label{fig:wide} 
(a) Simplified structural diagram of triplet ground state of $\rm V_B^-$. Relaxation between $\left|m_s = 0\right\rangle$ and $\left|m_s = \pm 1\right\rangle$ ($\left|m_s = -1\right\rangle$ and $\left|m_s = +1\right\rangle$) occurs at rate $\Omega$ ($\gamma$). (b) Optical and microwave pulse sequences used to extract the relaxation rates $\Omega$ (left) and $\gamma$ (right). 
(c) Normalized fluorescence decay data and single-exponential fits with decay rate $3\Omega$ (yellow line) and $2\gamma + \Omega$ (blue line).}
\end{figure*}

\section{Results and Discussion}

A silicon wafer with 280-nm thermally-grown SiO$_2$ on top is used as substrate. Developed by S1813 photoresist and ultraviolet lithography (SUSS MABA6) processes on the substrate, a 100-nm-thick Au coplanar waveguide (CPW) is deposited by LAB18 E-Beam Evaporator to radiate the microwave (MW) \cite{Gao2021High}. The flake hBN samples are mechanically exfoliated from the bulk-crystal hBN (HQ Graphene), and transferred onto the CPW of the substrate. 
We then use a helium ion microscope (HIM) to create a 3 $\times$ 3 $\rm V_B^-$ spin defect array, with an implantation dose of $10^{17}$ ions/$\rm cm^2$ and the implantation energy of 30 keV \cite{Zeng2023Reflective}. A large number of high-energy He$^+$ ions break the B-N bonds and bombard away boron atoms, leaving $\rm V_B^-$ spin defects when extra electrons are trapped in the boron vacancies. The schematic of the $\rm V_B^-$ defect in hBN is shown in Fig. 1(a). Illuminated by a 532-nm continuous wave excitation laser (a power of 0.8 mW) and filtered by a 750-nm long-pass filter, we get a confocal photoluminescence (PL) map, as shown in Fig. 1(b). 
Characteristic wavelength in PL spectrum of $\rm V_B^-$ defects ensemble defects at room temperature in Fig.1 (c) showing a broad profile centered around 810 nm and optically detected magnetic resonance (ODMR) spectrum measured at 0 mT and room temperature with theoretical two-Lorentzian fitting (red line) with frequencies $\nu_1$ = 3401.82 $\pm$ 1.88 MHz, $\nu_2$ = 3549.27 $\pm$ 1.89 MHz and $\nu_0 = (\nu_1 +\nu_2)/2 = $ 3475.55 $\pm$ 1.89 MHz, in Fig.1 (d) collectively identify the presence of $\rm V_B^-$ defects in hBN. 

Fig. 2(a) displays a simplified level structure of triplet ground state of $\rm V_B^-$ defect, with single-quantum relaxation between $\left|m_s = 0\right\rangle$ and $\left|m_s = \pm 1\right\rangle$ occurs at rate $\Omega$, and double-quantum relaxation between $\left|m_s = -1\right\rangle$ and $\left|m_s = +1\right\rangle$ at rate $\gamma$, respectively \cite{Cambria2023Temperature,Myers2017Double}. 
To extract the single-quantum rate $\Omega$ and double-quantum rate $\gamma$, we perform two sets of measurements using the two sets of optical and microwave pulse sequences shown in Fig. 2(b) \cite{Myers2017Double}. 
We first polarize the $\rm V_B^-$ spins to $\left|m_s = 0\right\rangle$ state with a long green laser pulse, and then let the spins freely evolve $\tau$ time, and finally read out the fluorescence count ($S_{0,0}(\tau)$) representing the spin state with another short green laser pulse. After a similar sequence that adds a $\pi$-pulse whose frequency is resonant with the spin-level splitting between $\left|m_s = 0\right\rangle$ and $\left|m_s = -1\right\rangle$ ($\pi_{0,-1}$) before readout, we get $S_{0,-1}(\tau)$. The $\pi_{0,-1}$-pulse flips the population of $\left|m_s = 0\right\rangle$ and $\left|m_s = -1\right\rangle$, so the $F_1(\tau)=S_{0,0}(\tau)-S_{0,-1}(\tau)$ reflects the population difference of $\left|m_s = 0\right\rangle$ and $\left|m_s = -1\right\rangle$.
Based on the solution of the rate equations governing the populations of the three sublevels, $F_1(\tau)$ is proportional to $e^{-3\Omega \tau}$ (see Section 2 in Supporting Information for detailed derivation process).
Similarly, for the right sequence in Fig. 2(b), the $\rm V_B^-$ spins are initialized to $\left|m_s = -1\right\rangle$ after a polarized laser pulse and a $\pi_{0,-1}$-pulse and then freely evolve $\tau$ time, and finally the fluorescence count $S_{-1,-1}(\tau)$ ($S_{-1,+1}(\tau)$) is read out after a $\pi_{0,-1}$ ($\pi_{0,+1}$)-pulse. The $F_2(\tau)=S_{-1,-1}(\tau)-S_{-1,+1}(\tau)$ reflects the population difference of $\left|m_s = -1\right\rangle$ and $\left|m_s = +1\right\rangle$, and is proportional to $e^{-(2\gamma+\Omega)\tau}$. 
Fig. 2(c) displays the experimental and fit results at room temperature (293 K), giving that $\Omega =$ 33.26 $\pm$ 1.83  kHz and $\gamma =$ 81.60 $\pm$ 9.12 kHz. 
The values of $\gamma$ exceeds that of nitrogen vacancy (NV) in diamond by almost four orders of magnitude, and we attribute this phenomenon to the presence of larger number of phonon mode energies in hBN, which leads to more complex spin-phonon interactions \cite{Cambria2023Temperature}. 

\begin{figure*}
\includegraphics[width=\textwidth,height=0.65\textwidth]{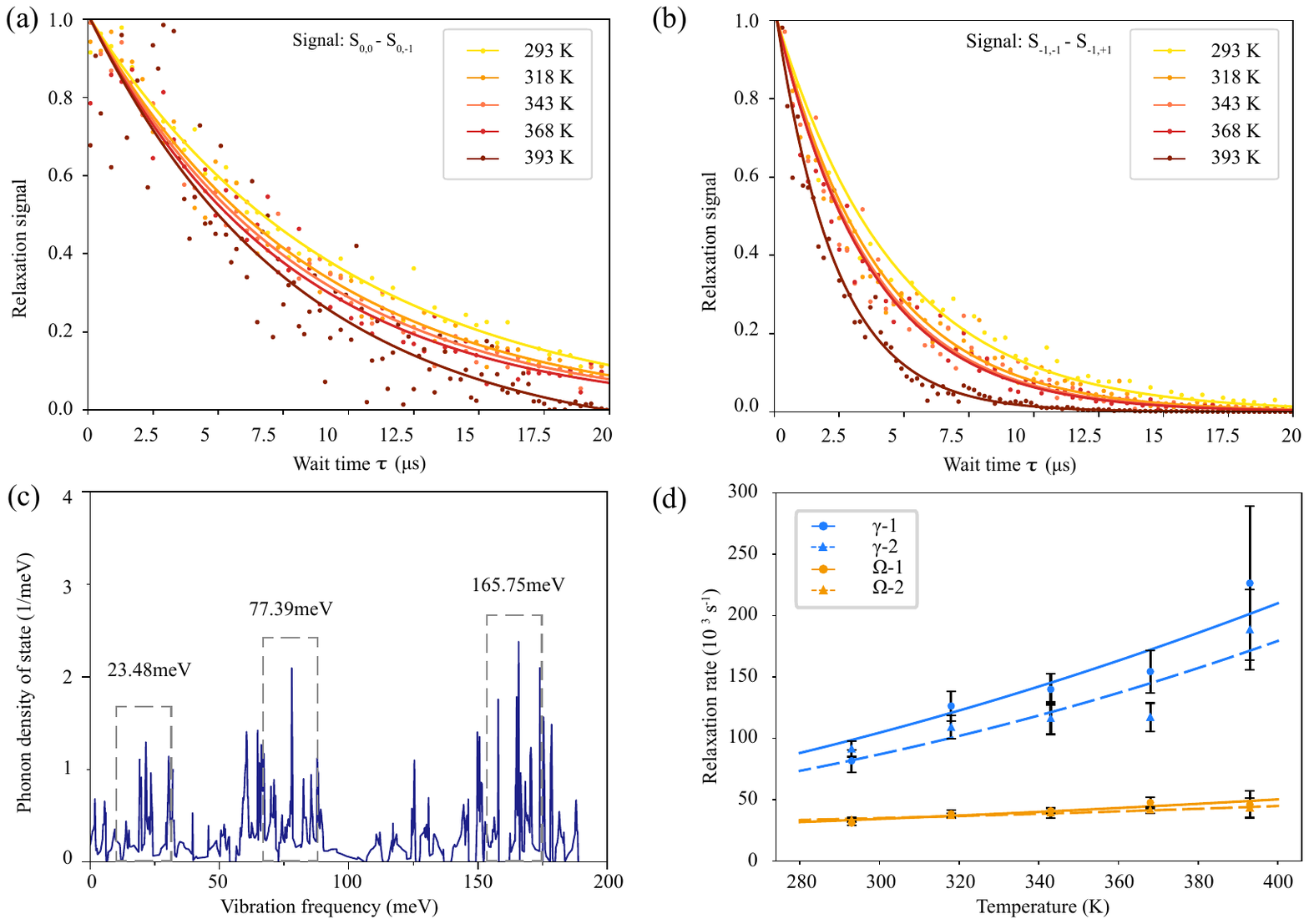}
\caption{\label{fig:wide}(a) Temperature dependent normalized fluorescence decay data ($F_1(\tau)=S_{0,0}(\tau)-S_{0,-1}(\tau)$), and single-exponential fits with decay rate $3\Omega$.
(b) Temperature dependent normalized fluorescence decay data ($F_2(\tau)=S_{-1,-1}(\tau)-S_{-1,+1}(\tau)$), and single-exponential fits with decay rate $2\gamma+\Omega$.
(c) Phonon density of states (PDOS) obtained from first-principle calcutations as a function of phonon frequencies of the defective 3 $\times$ 3 monolayer hBN, peaked at 23.48, 77.39 and 165.75 meV, regarded as the dominant characteristic phonon modes contributing phonon-spin relaxation process.
(d) Temperature dependence of relaxation rate $\Omega$ (yellow dots) and $\gamma$ (blue dots) of Spot 1 (circle) and Spot 2 (triangle), fitted with equations (2) (yellow lines) and (3) (blue lines).} 
\end{figure*}

Furthermore, we measure the single- and double-quantum relaxation rates of Spot 1 and Spot 2 marked with white dotted circles in Fig. 1(b) at various temperatures from 293 to 393 K. A lab-built temperature control system which can achieve stability within $\pm$ 100 mK is used (see Section 1 in Supporting Information for details).  
The experimental results of Spot 2 (Spot 1) are displayed in Fig. 3(a) and (b) (Fig. S1(e) and (f) in Supporting Information), and the relaxation rates $\Omega$ and $\gamma$ of varying temperatures obtained by fit are displayed in Fig. 3(d). Both rates increase with increasing temperature, and the double-quantum relaxation rate $\gamma$ significantly increases rapidly. At 400 K, $\gamma$ can reach the values five times of $\Omega$.
The total spin-lattice relaxation time $T_1$ that considers both $\Omega$ and $\gamma$ and limits coherence time $T_2$ should be\cite{Myers2017Double}
\begin{equation}
    \frac{1}{T_1}=3\Omega+\gamma
\end{equation}
Our results indicate that double-quantum relaxation rate dominates the decoherence channel of spin-phonon interactions at high temperature. 

The temperature dependence of the single- and double-quantum relaxation rate is mainly attributed to spin-phonon interactions. Prior work has developed a simplified theoretical model in which second-order spin-phonon interactions play a leading role, and successfully explained the interaction between NV spin and phonon \cite{Cambria2023Temperature}.
Analogize to $\rm V_B^-$ spins and based on the triple-peak form of the phonon vibration spectral function \cite{Liu20225temperature}, the relaxation rates may be approximately expressed as
\begin{equation}
    \Omega(T) = \sum_{i=1,2,3}A_in_i(n_i+1) + A_S
\end{equation}
\begin{equation}
    \gamma(T) = \sum_{i=1,2,3}B_in_i(n_i+1) + B_S
\end{equation}
where $n_{i}=(e^{\hbar\omega_i/{k_BT}}-1)^{-1}$ (i=1,2,3) are the mean occupation numbers of the phonon modes, $A_{i}$ and $B_{i}$ are coupling coefficients associated with the effective modes, and $A_S$ and $B_S$ are sample-related constants. 
Using density functional theory (DFT) \cite{Kresse1996Efficiency,Kresse1996Efficient} with Perdew-Burke-Ernzerhof (PBE) \cite{Perdew1996Generalized} functional, we perform first-principle simulations. The phonon density of states (PDOS), as a function of phonon frequencies, is shown in Fig. 3(c). The spectral function displays three peaks around 23.48, 77.39 and 165.75 meV respectively \cite{Liu20225temperature,Estaji2025spin}. 
By substituting these three characteristic phonon energies into Equation (2) and (3), we fit the experimental data corresponding to $\Omega$ and $\gamma$ and good fit was achieved (fitting curves in Fig. 3(d)). 
In addition, the fitting results show that the higher energy phonon mode, the greater the coupling coefficient (see Section 4 in Supporting Information), which is consistent with that of NV centers in diamond previously reported. 

\section{Conclusion}

To summarize, we measure the temperature dependent single- and double-quantum spin-lattice relaxation process of $\rm V_B^-$ defects. 
Both single- and double-quantum relaxation rates increase when temperature increases. Double-quantum relaxation rate $\gamma$ increases more rapidly and can reach the value five times that of single-quantum relaxation rate $\Omega$ at 400 K. 
We attribute the high-temperature behaviour of spin-lattice relaxation rates of $\rm V_B^-$ defect to second-order interactions with higher-energy effective phonon mode.
The double-quantum relaxation may dominate the decoherence channel of spin-phonon interactions, holding promise for extension to other spin defects in hBN. 

\section*{Acknowledgments}

This work is supported by the Innovation Program for Quantum Science and Technology (No. 2021ZD0301200), the National Natural Science Foundation of China (Nos. 12174370, 12174376, 11821404, 12304546, 124B2082 and 62304161), the Youth Innovation Promotion Association of Chinese Academy of Sciences (No. 2017492), Anhui Provincial Natural Science Foundation (No.2308085QA28), China Postdoctoral Science Foundation (No. 2023M733412). This work was partially carried out at the USTC Center for Micro and Nanoscale Research and Fabrication.

\end{document}